\documentstyle[twoside,fleqn,espcrc2,psfig]{article}

\newcommand{\NP}[1]{ Nucl.\ Phys.\ {\bf #1}}

\newcommand{\PL}[1]{ Phys.\ Lett.\ {\bf #1}}

\newcommand{\PR}[1]{Phys.\ Rev.\ {\bf #1}}
\newcommand{\PRL}[1]{ Phys.\ Rev.\ Lett.\ {\bf #1}}

\newcommand{\AmS}{{\protect\the\textfont2
  A\kern-.1667em\lower.5ex\hbox{M}\kern-.125emS}}

\title{Thermal evolution of the chiral condensate in 
SU(2) and SU(3) Chiral Perturbation Theory. }    
\author{
J.R.Pel\'aez
\address{
Universit\`a degli Studi di Firenze and  I.N.F.N. Sezione di Firenze, Italy
\\and\\
Universidad Complutense de Madrid.
28040 Madrid. Spain}
}

\begin{document}

\begin{abstract}
The temperature evolution of the chiral condensates
 in a gas made of pions, kaons and etas
is studied within the framework of SU(2) and SU(3) Chiral Perturbation Theory.
We describe the temperature dependence of the
quark condensates
by using the meson meson scattering phase shifts
in a second order virial expansion. 
We find that the SU(3) formalism yields
an extrapolated melting temperature for the  non-strange condensates
which is lower by about 20-30 MeV than within SU(2).
In addition our results show that the strange condensate melting is 
slower than that of the non-strange, due to the 
different strange and non-strange quark masses.
\end{abstract}

\maketitle

\pagestyle{empty}
\arraycolsep=0.5pt
\section{Introduction}

We review here our recent study of the thermal
evolution of the chiral strange and non-strange condensates \cite{yo}.
Our motivation is the study of the 
the phase diagram of QCD at zero baryonic densities, which is 
a state that is expected to occur in the central rapidity regions
of Relativistic Heavy Ion Collisions.
Our purpose  is to obtain a model independent description 
of the condensate and its thermal evolution. 
To that aim we will turn to the virial expansion and to  Chiral 
Perturbation Theory.

The virial expansion is a simple and successful approach to describe
many thermodynamic features of dilute gases made of interacting pions
\cite{virial2} and other hadrons \cite{virial3}. 
Let us note that for most thermodynamic properties it is enough to know
the low energy scattering phase shifts of the particles,
which, in principle,  could be taken from experiment, avoiding
any model dependence.
However, we are interested in the
chiral symmetry restoration, whose order parameters are the
quark condensates, which are defined as derivatives of the 
pressure with respect to the quark masses. These derivatives cannot be 
measured experimentally
 and therefore a model independent
theoretical description of the 
scattering amplitudes mass dependence  is required. 
That is why we need the second ingredient of our approach:
Chiral Perturbation Theory (ChPT) \cite{Weinberg,chpt1,GL3},
which
provides a model-independent and systematic description of 
low energy hadronic interactions. 
ChPT follows from the identification of the pions, kaons
and the eta as the Goldstone Bosons of the QCD spontaneous
chiral symmetry breaking (pseudo Goldstone bosons indeed, due to the
small light-quark masses). Then,  the ChPT Lagrangian is built as
the most general 
derivative and mass expansion, over $4\pi F\simeq 1.2\, \hbox{GeV}$,
(the symmetry breaking scale) respecting the
symmetry constraints. At one loop any calculation can be renormalized
in terms of just an small set of parameters, $L_k(\mu)$,
 $H_k(\mu)$, ($\mu$ being the renormalization scale)
which can be determined from a few experiments 
and used for further predictions.

In particular, within SU(2) ChPT 
the leading
coefficients of the low temperature expansion for the pressure 
and the non-strange quark condensate have
been calculated for
a hadron gas whose only interacting hadrons were the pions \cite{Gerber}. 
Adding other more massive particles
(like kaons, etas, rhos, protons, etc...)  as {\it free states},
it was possible to obtain an estimate, $T\simeq190-200\,$MeV,
of the phase transition critical temperature.  
It was also shown that the perturbative
calculation of the pressure was analogous
to the second order virial expansion when the interacting
part of the second virial coefficient is obtained from
 the one loop $\pi\pi$ ChPT scattering lengths. Other works
have studied the applicability of the virial expansion with
a pionic chemical potential \cite{nos}, 
or the critical temperature in generalized
scenarios \cite{GChPT}.

The interest of extending the previously commented work to SU(3)
is that the SU(2) approach is limited by the absence of
other interacting particles, like kaons or etas, whose
densities at $T\simeq150$ MeV are significant \cite{Gerber}. 
In addition, let us recall that
the chiral phase transition can be different 
for the SU(2) and SU(3) cases, and that
several QCD inequalities and lattice results suggest an 
stronger
chiral condensate temperature suppression 
with an increasing number of light flavors \cite{flavors}.
Intuitively this is due to the fact that the existence of other states
(strange quarks, or  strange mesons)
makes it easier to create entropy, that is, disorder,
and therefore it is easier to melt the ordered state,
that is, the condensates.
This effect  has been observed in lattice calculations \cite{lattice},
but only in the chiral limit, and it seems to lower
the chiral critical temperature down by roughly
$20\;$ MeV. Note, however,  
that none of these results have been obtained from the
hadronic phase and realistic masses, which is what we have 
achieved with our approach.

In addition, within SU(2) it is not possible to study the 
$\langle\bar{s}s\rangle$ condensate. This could be interesting
because, in the 
chiral phase diagram, 
quark masses play  the same role as magnetic fields
in ferromagnets: Intuitively, we need 
a higher temperature
to disorder a ferromagnet when there is a magnetic field
aligned along the direction of the magnetization.
Analogously, it has been found that the SU(2) chiral condensate
melts at a lower temperature in the massless limit \cite{Gerber}.
In particular, we have obtained
the temperature dependence of $\langle\bar{s}s\rangle$
in the hadronic phase. Since the strange quark mass 
$m_s$ is much larger
than $m_u, m_d$, 
we have indeed found a sizable ``ferromagnetic'' effect which translates
into a slower
thermal $\langle\bar{s}s\rangle$ evolution. This effect had already been studied in the
large $N_c$ limit with the use of QCD-motivated effective Lagrangian 
\cite{Hatsuda}.

Thus, the second order relativistic virial expansion of the pressure
for   an inhomogeneous gas made of
three species: $i=\pi,K,\eta$ is \cite{Dashen,virial2}:
\begin{equation}
\beta P=\sum_i B_{i}(T)\xi_i + 
\sum_i\left( B_{ii}\xi_i^2 + \frac{1}{2}\sum_{j\neq i}
B_{ij}\xi_i\xi_j
\right)...,
\nonumber
\end{equation}
where $\beta=1/T$ and $\xi_i=\exp({-\beta m_i})$. Expanding
up to the second order in $\xi_i$ means that we consider
only binary interactions. For a free boson gas, 
$B_{ij}^{(0)}=0$ for $i\neq j$, whereas:
\begin{eqnarray}
B_i^{(0)}&=& \frac{g_i}{2\pi^2}\int_0^{\infty} dp\, 
p^2 e^{ -\beta (\sqrt{p^2+m_i^2}- m_i)},\\
B_{ii}^{(0)}&=&\frac{g_i}{4\pi^2}\int_0^{\infty} dp\, 
p^2 e^{ -2\beta (\sqrt{p^2+m_i^2} -m_i)}.
\end{eqnarray} 
The degeneracy is $g_i=3,4,1$ for $\pi, K, \eta$, respectively.
The interactions appear through \cite{Dashen,Gerber}:
\begin{eqnarray}
B_{ij}^{(int)}=\frac{\xi_i^{-1}\xi_j^{-1}}{2\,\pi^3}\int_{m_i+m_j}^{\infty} dE\, E^2 K_1(E/T) 
\Delta^{ij}(E),\nonumber
\end{eqnarray}
where $K_1$ is the first modified Bessel function
and:
\begin{equation}
\Delta^{ij}=\sum_{I,J,S} (2I+1)(2J+1)\delta^{ij}_{I,J,S}(E),
\label{Delta}
\end{equation}
$\delta^{ij}_{I,J,S}$ being the $ij\rightarrow ij$
phase shifts (chosen so that $\delta=0$ at threshold) 
of the  elastic scattering of a state
$ij$ with quantum numbers $I,J,S$ ($J$ being the total angular momentum
and $S$ the strangeness).

The virial expansion breaks around
$T\simeq 200-250$ MeV \cite{nos}. Fortunately,
the physics we are interested in occurs already at  $T<250\,$MeV,
and $\xi_\pi> \xi_K \simeq \xi_\eta$, and it is enough
to consider $ij=\pi\pi,\pi K$ and $\pi \eta$ 
in $B_{ij}$.

At this point we  recall that
the quark mass appears in the QCD Lagrangian, 
and therefore in the partition function, as 
$m_{q_\alpha} \bar{q}_\alpha q_\alpha$,
where $q_\alpha=u, d,s$.
Thus, in order to obtain the condensate we simply have to 
differentiate the partition function with respect to $m_{q_\alpha}$.
At finite temperature, the partition function 
is substituted by the free energy density $z$,
so that \cite{Gerber}
\begin{equation}
\langle\bar{q}_\alpha q_\alpha\rangle=\frac{\partial z}{\partial m_{q_\alpha}}
=\langle 0 \vert\bar{q}_\alpha q_\alpha\vert 0 \rangle-  
\frac{\partial P}{\partial m_{q_\alpha}}.
\label{condmq}
\end{equation}
Note that we have separated the $T=0$ part from the 
temperature dependent part, which is nothing but the pressure
$P=\epsilon_0- z$, $\epsilon_0$ being the  vacuum energy density.
Of course, at $T=0$ we recover 
$\langle\bar{q}_\alpha q_\alpha\rangle=\langle 0 \vert\bar{q}_\alpha q_\alpha\vert 0 \rangle=\partial \epsilon_0/\partial m_{q_\alpha}$. 

Once more, we  emphasize that
in contrast with most thermodynamic quantities, for
the chiral condensate it is not enough
to know the $\delta(E)$, but {\it we also need
their dependence with the quark masses} as well as a value
for the $T=0$ vacuum expectation value. We will  obtain
this information  from
ChPT. Nevertheless, ChPT does not deal with quarks,
and thus we first have to 
rewrite the condensate, eq.(\ref{condmq}), in terms of meson masses:
\begin{equation}
\langle\bar{q}_\alpha q_\alpha\rangle
=\langle 0 \vert\bar{q}_\alpha q_\alpha\vert 0 \rangle 
\left(1+\sum_i \frac{c^{\bar{q_\alpha}q_\alpha}_i}{2 m_i F^2} 
\frac{\partial P}{\partial m_i}\right)
\end{equation}
where, as before, $i=\pi,K, \eta$, and we have defined:
\begin{equation}
c^{\bar{q_\alpha}q_\alpha}_i=- F^2\frac{\partial m_i^2}
{\partial m_{q_\alpha}}\langle 0 
\vert\bar{q}_\alpha q_\alpha\vert 0 \rangle ^{-1}.
\end{equation}
In the isospin limit,
when $m_u=m_d$, the $u$ and $d$ condensates
are equal, and we define
$\langle 0\vert\bar{q} q\vert 0 \rangle \equiv\langle 0\vert\bar{u} u+\bar{d} d\vert 0 \rangle$.
It is tedious but straightforward to obtain the $c$ parameters above, 
whose explicit expressions can be found in \cite{yo}.
The only relevant comment is that the $c$ coefficients depend on
the chiral parameters $L_k$, for $k=4...8$, and $H_2$.

In Table I, we show two $L_k$ determinations
from meson data, but  it makes little difference to use
other parameter sets. For $H_2$ we will
use $H_2^r(M_\rho)=(-3.4\pm1.1)10^{-3}$, obtained 
as explained in \cite{Jamin} but using a more
recent estimation of 
$\langle 0\vert\bar{s}s\vert 0\rangle/\langle 0\vert\bar{q}q\vert 0\rangle=0.75\pm 0.12$
\cite{Narison}.
For instance,  the
$c$ parameters obtained when using the $L_k$ in the first
column of Table I are:
\begin{eqnarray}
c^{\bar{q}q}_\pi=0.9^{+0.2}_{-0.4} , c^{\bar{q}q}_K=0.5^{+0.4}_{-0.7},
c^{\bar{q}q}_\eta=0.4^{+0.5}_{-0.7}, \nonumber\\
c^{\bar{s}s}_\pi=-0.005^{+0.029}_{-0.037}, c^{\bar{s}s}_K=1.3^{+0.4}_{-0.8},
c^{\bar{s}s}_\eta=1.5^{+0.9}_{-1.6}. \nonumber
\end{eqnarray}
The above  $c^{\bar{q}q}_\pi$ is in good
agreement with the SU(2) estimates: $0.85$ and  $0.90\pm0.05$ \cite{Gerber}.
\begin{table}
\begin{tabular}{|c||c|c|}
\hline
$\mu=M_\rho$&Refs.\cite{chpt1,BijnensGasser}
&IAM \cite{angelyo}
\\ \hline
$L_1^r$
& $0.4\pm0.3$
&$0.561\pm0.008\,(\pm0.10)$\\
$L_2^r$
& $1.35\pm0.3$
&$1.21\pm0.001\,(\pm0.10)$\\
$L_3$
& $-3.5\pm1.1$
&$-2.79\pm0.02\,(\pm0.12)$\\
$L_4^r$
& $-0.3\pm0.5$
&$-0.36\pm0.02\,(\pm0.17)$\\
$L_5^r$
& $1.4\pm0.5$
&$1.4\pm0.02\,(\pm0.5)$\\
$L_6^r$
& $-0.2\pm0.3$
&$0.07\pm0.03\,(\pm0.08)$\\
$L_7$
& $-0.4\pm0.2$ 
&$-0.44\pm0.003\,(\pm0.15)$\\
$L_8^r$
& $0.9\pm0.3$ 
&$0.78\pm0.02\,(\pm0.18)$\\
\hline
\end{tabular}
\caption{\rm Different sets of chiral parameters $\times10^{3}$. 
In the first column $L^r_1,L^r_2,L_3$ are taken from
\cite{BijnensGasser}
and  the rest from  \cite{chpt1}
($L^r_4$ and $L^r_6$ are estimated from the Zweig rule).
The last column is the IAM fit \cite{angelyo}
to meson-meson scattering up to 1.2 GeV.
}
\label{eleschpt}
\end{table}

Finally, we  have to introduce the meson-meson interactions
in the second order virial coefficients through the phase shifts in
eq.(\ref{Delta}). These are nothing but the complex phases
of the meson-meson amplitudes, once they are projected 
in partial waves of definite isospin and angular momentum.
The one loop ChPT amplitudes have been given in
\cite{o4,angelyo}. As already commented
within SU(2) ChPT it has been shown \cite{Gerber}
 that using just the amplitudes at threshold 
(scattering lengths) 
in the virial expansion is equivalent to expanding the partition
function to third order in $T/F$ or $T/m_\pi$. This approach yields
a fairly good representation of the pion gas thermodynamics
at low temperatures, $T\ll 150\,$MeV \cite{Gerber}. Let us recall
that ChPT provides  a good low energy description ($E<500\,$MeV)
of the meson-meson amplitudes \cite{o4}.

The virial coefficients and 
$\partial P/\partial m_i$ have been calculated numerically.
For simplicity, in the figures, we have represented the 
chiral condensate over its vacuum expectation value,
$\langle \bar{q}_\alpha q_\alpha
\rangle /\langle 0 \vert\bar{q}_\alpha q_\alpha\vert 0 \rangle $,
so that all of them are normalized to 1 at $T=0$.
As a matter of fact, since there is always an small explicit chiral symmetry 
breaking due to the quark masses, the condensate should only 
vanish completely in the $T\rightarrow\infty$ limit
(following with our analogy between the quark masses and the magnetic field, our ferromagnet above
the Curie point becomes paramagnetic, but as if it was 
 in the presence of a magnetic field, which still
produces some magnetization.).
Of course, with a virial expansion truncated at second order we cannot 
generate such an analytic behavior, and our curves become negative
above some $T$, where the approach becomes clearly unreliable. 
Nevertheless, since the $u$ and $d$ masses are so small compared
with the size of the non-strange condensate at T=0 (less than 3\%),
it is a fairly good approximation to speak about a melting temperature
in that case. For the $\langle\bar{s}s\rangle$ condensate,
we will only consider an extrapolated  melting temperature to ease
the comparison with the non-strange case, but that number 
does not have an actual physical meaning.

\begin{figure}
\psfig{file=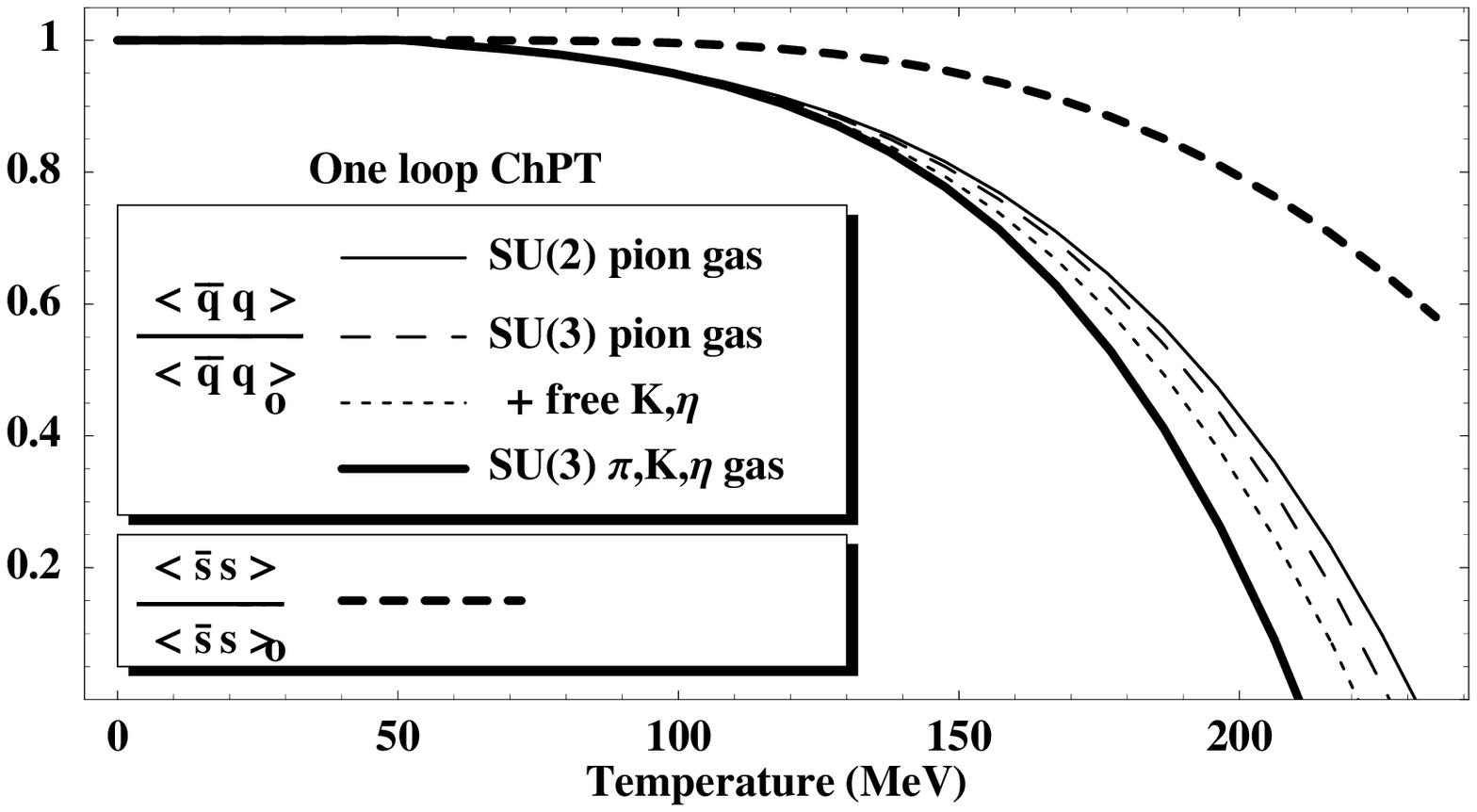,width=7.6cm}
{\footnotesize {\bf Figure 1.}
\rm Condensate thermal evolution using one-loop ChPT
amplitudes. Although {\it they should not really vanish}
at a finite temperature,
they are extrapolated down to zero only for reference.}
\end{figure}

In Figure 1 we show the results of using the 
one-loop ChPT  phase shifts obtained using
the central values of the parameters in the first column of Table I.
For illustration, we have separated the different contributions. 
In particular, 
the thin continuous line represents the interacting pion gas in SU(2),
(see \cite{GL3} for the conversion of SU(3) 
into SU(2) parameters), whereas
the thin-dashed line is the corresponding
one-loop SU(3) result also for a pion gas. We have checked that
the tiny difference between them comes only from the pure $O(p^4)$
phase shift contribution, ( the amplitude, once the parameters are translated,
is  the same for SU(2) and SU(3) only up to $s/M_K^2$ or $s/M_\eta^2$
terms coming from the expansion of kaon or eta loops, which,
of course, are not present in SU(2) \cite{GL3}). 
This amounts to a  4 MeV decrease
of the pion gas extrapolated melting temperature: 
$T_m^{\langle \bar{q} q\rangle}=231\,$MeV for SU(2) and 
$T_m^{\langle \bar{q} q\rangle}=227\,$MeV for SU(3). 
Next, the thin-dotted line is obtained 
by adding free kaons and etas to the pion gas,
which lowers the extrapolated melting temperature, by roughly
5-6 MeV down to 221 MeV.

Our SU(3) result for a 
$\pi, K, \eta$ gas, where also the $kaons$ and $etas$ interact with 
each other and the pions, is the thick continuous line,
where we see that the additional decrease, basically  due to the
$\pi K,\pi\eta$ interactions, amounts to 10-11 MeV.
On a first glance, it may seem surprising that the 
$\pi K, \pi\eta$ interaction effect comes out comparable
or larger than that of  free kaons and etas, since it
is thermally suppressed by another $\exp(-m_\pi/T)$ factor. 
Note, however, that the $\exp(-m_\pi/T)$ suppression
only amounts to a factor of $6,4,2.5$ 
at T= 80,100,150 MeV, respectively.
In contrast, the $\pi K$ 
and $\pi\eta$ interactions depend strongly on $m_\pi$, 
which is much more
sensitive to $\hat m$ than  $m_K$ or $m_\eta$, which 
carry the 
only  $\hat m$ dependence of the free $K$ and $\eta$ terms.
In particular, it is easy to see that
$\partial m_\pi^2/\partial \hat m\simeq 2 \, \partial m_K^2/\partial\hat m$
and
$\partial m_\pi/\partial \hat m\simeq 6 \, \partial m_K/\partial\hat m$, 
and this ``temperature independent enhancement'' 
competes with the thermal suppression, making the $\pi K$ and $\pi \eta$
interaction effect comparable to the free one, already at
$T=80-100\,$MeV and even larger if $T>140\,$MeV.

Putting all the pieces together, 
the $T_m^{\langle \bar{q} q\rangle}$ decrease in SU(3) is
roughly 20 MeV, in agreement 
with the chiral limit lattice 
results quoted in  \cite{lattice}:
 $T_c=173\,$MeV for SU(2) and  $T_c=154\,$MeV,
for SU(3). The effect of using the real masses
would be to increase both temperatures.

Furthermore,
 it can be noticed that $\langle \bar{s} s\rangle$ melts much slower
than  $\langle \bar{q} q\rangle$,
since $m_s\gg \hat m$. Indeed, there is still 70\% left of
the $\langle \bar{s} s\rangle$ condensate at the  $\langle \bar{q} q\rangle$
melting point. 

Let us recall that both effects are already sizable at 
low temperatures $T\simeq100\, $MeV, where we can still trust 
pure ChPT. As commented above, only  for illustration and to ease 
the comparison between curves and with previous works,
we have extrapolated the condensates down to zero.

At this point we can also consider the 
uncertainties in the chiral parameters in Table I.
By performing a Montecarlo gaussian sampling of the
parameters within their errors, we can obtain a 
rather {\it conservative} estimate of our uncertainties
(since the sampling is done assuming uncorrelated errors).
We thus find:
\begin{eqnarray*}
T_m^{\langle \bar{q} q\rangle}
= 211^{+19}_{- 7}\,\hbox{MeV}, \;
T_m^{\langle \bar{s} s\rangle}
= 291^{+37}_{-35}\,\hbox{MeV}
\end{eqnarray*}
Note that the $\langle \bar{s} s\rangle$ extrapolation is 
beyond the reliability region of the second order
virial expansion, is just illustrative, and really only vanishes in the 
$T\rightarrow\infty$ limit.
Let us also remark that the  extrapolated melting temperatures 
are strongly correlated, so that their difference
has less uncertainty than simply adding in quadrature
their respective errors. In particular, we find
\begin{eqnarray*}
&&T_m^{\langle \bar{q}q\rangle, SU(2)}-
T_m^{\langle \bar{q}q\rangle, SU(3)}= 
21^{+14}_{-7}\,\hbox{MeV}
\\
&&T_m^{\langle \bar{s}s\rangle}-
T_m^{\langle \bar{q}q\rangle}= 
80^{+25}_{-40}\,\hbox{MeV}
\end{eqnarray*}
We remark once more that 
the $\langle \bar{s}s\rangle$ melting temperature
is, of course, only a crude extrapolation for illustrative purposes
and to ease the comparison with the non-strange condensate
evolution.

Furthermore, we have estimated the effect of other, 
more massive, hadrons. In the SU(2) case \cite{Gerber,nos}
these states also included the kaons and etas, although
assuming a rather large uncertainty in 
$\partial M_i/\partial m_{q_\alpha}$. All in all this effect
reduced $T_m^{\langle \bar{q}q\rangle}$ by approximately
10-20 MeV. In the SU(3) case we have indeed calculated
explicitly the kaon and eta $\hat m$ dependence
in the $c^{\bar{q_\alpha}q_\alpha}$ parameters, thus
reducing considerably the above uncertainty. 
Since the kaons and eta are treated explicitly,
the other massive hadrons are heavier than $m_\eta$, have a 
very low density, and their main contribution to the pressure
comes from the first virial coefficient, i.e. the free gas.
The only uncertainty is on $\partial M_h/\partial m_{q_\alpha}$, 
conservatively estimated to lie within
the number of valence quarks $N_{q_\alpha}$ and 
$2N_{q_\alpha}$. Thus, we have found that their 
contribution decreases $T_m^{\langle \bar{q} q\rangle}$
by 7-12 MeV, respectively.

Finally, and in order to estimate the effects of higher energies
we can extend ChPT
up to $E\simeq1.2\,$GeV by means of unitarization models
\cite{IAM1,IAM2,angelyo}.
These techniques resum the ChPT series respecting unitarity but
also the low energy expansion, {\it including the mass dependency}.
In particular, it has been shown that the coupled channel
Inverse amplitude Method (IAM) provides a remarkable and accurate description
of the complete meson-meson interactions 
 below 1.2 GeV, generating dynamically
six resonances  from
the one-loop ChPT expansion and a set of fitted $L_k$ compatible
with other previous determinations. This approach can be 
extended systematically to higher orders 
(for SU(2) case, see \cite{Juan}).

Hence, in Figure 2 we show the results of using the
one-loop coupled channel IAM fitted phase shifts. Its
corresponding $L_i$ parameters are given in
the last column of Table I, with two errors, the first, very small, is
purely statistical, and the second covers the uncertainty in the
parameters depending on what systematic error is assumed for 
the experimental data. Let us remark that, although 
 the $L_k$ are highly correlated, the second, larger error, 
ignores completely these correlations, 
and should be considered as a very conservative range.   
The continuous line corresponds to the central values, 
and the dark shaded areas cover the one standard deviation uncertainty
due to the small errors in the parameters.
These areas have been obtained from a Montecarlo Gaussian sampling.
The conservative ranges are covered by the light gray areas.
We now find:
\begin{eqnarray*}
T_m^{\langle \bar{q} q\rangle}
= 204^{+3}_{-1}\,\,(^{13}_{\,\,5})\, \hbox{MeV},
T_m^{\langle \bar{s} s\rangle}
= 304^{+39}_{-25}\,\,(^{120}_{\,\,65})\, \hbox{MeV}
\end{eqnarray*}
where the errors in parenthesis are the conservative
errors which should be interpreted
as uncertainty ranges better than as  standard deviations. 
Note the excellent agreement with standard ChPT. 
The magnitude of the different contributions is roughly the same,
although the $\pi K$ and $\pi\eta$ interactions this time lower
$T_m^{\langle \bar{q} q\rangle}$ by 17 MeV, 
and the free kaons and etas by roughly 10 MeV.
Again other massive states apart from kaons and etas
would lower $T_m^{\langle \bar{q} q\rangle}$
by 6-10 MeV, and $T_m^{\langle \bar{s} s\rangle}$ by 22-32 MeV.
In addition, we find
 \begin{eqnarray*}
&&\Delta T_m=T_m^{\langle \bar{q}q\rangle SU(2)}-
T_m^{\langle \bar{q}q\rangle SU(3)}= 
31.50^{+1.20}_{-0.03}\,\,(^{9}_{\,\,8})\,\hbox{MeV}
\\
&&\Delta T_m=T_m^{\langle \bar{s}s\rangle}-
T_m^{\langle \bar{q}q\rangle}= 
100^{+36}_{-29}\,\,(^{120}_{\,\,80})\,\hbox{MeV},
\end{eqnarray*}
The IAM results show that higher energy effects do not
affect very much our conclusions and that the ChPT extrapolation
from low temperatures seems robust.

\begin{figure}
\psfig{file=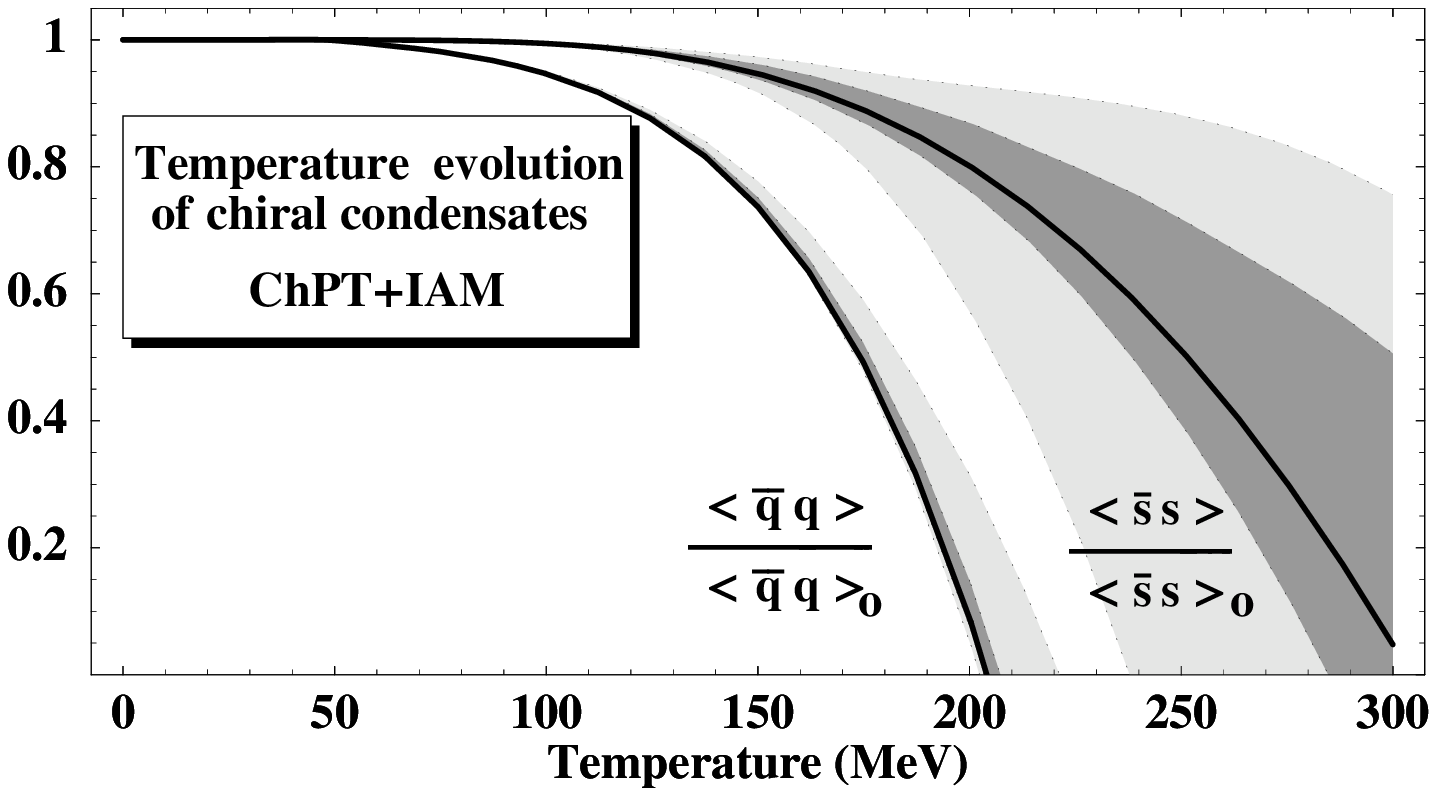,width=7.5cm}
{\footnotesize {\bf Figure 2}
\rm Temperature
evolution of chiral condensates using the unitarized ChPT amplitudes. 
The shaded areas cover the uncertainties
in different sets of chiral parameters. Although 
{\it they should not really vanish}
they are extrapolated down to zero only to ease their comparison 
and for reference.}
\end{figure}

To summarize, we have studied the SU(2) and SU(3) 
temperature evolution of the chiral condensates {\it in the
hadronic phase}. The thermodynamics of the meson gas has been 
obtained from the virial expansion and Chiral Perturbation Theory.
Our results clearly show a significant 
decrease, about 20-30 MeV,  of the non-strange condensate melting temperature,
from the SU(2) to the SU(3) case, similar to lattice results.
 Of these, about 6 MeV
had already been explained with free kaons and etas, but
the rest are mainly due to $\pi K$ and $\pi \eta$ interactions.
We have also estimated the effect of heavier hadrons and
of the third order virial coefficient. All in all
we find 
\begin{eqnarray*}
T_m^{\langle \bar{q} q\rangle}
= 201^{+23}_{-11}\,\hbox{MeV},
\end{eqnarray*}

In addition, our results show an slower 
temperature evolution of the strange condensate, shifted
by about 80 MeV with respect to the non-strange, 
due to the different quark masses.
More quantitatively, $\langle \bar{q} q\rangle$
does not show a sizable melting up to $T\simeq150\,$MeV and
 it still remains about 80\% when the non-strange is basically zero.

Both effects are clearly visible already at low temperatures.
However, we have checked that
their size is completely similar when using unitarized models.

These techniques should be easily extended 
to Heavy Baryon Chiral Perturbation Theory, in order to study
the condensates with non-zero baryon density. In general the whole
approach could be used with any effective Lagrangian formalism,
in particular to study other QCD phase transitions like those of the
color superconducting phases.

\vspace{.1cm}
 The author
thanks  A. Dobado, A. G\'omez Nicola and E. Oset 
for useful comments,
and support from the Spanish CICYT projects
FPA2000 0956 and BFM2000 1326,a CICYT-INFN collaboration grant,
the European EURIDICE HPRN-CT-2002-00311, as well as a
Marie Curie fellowship MCFI-2001-01155.
\appendix


\bibliography{apssamp}


\end{document}